\documentclass{webofc}
\usepackage[varg]{txfonts}   
\usepackage{graphicx}
\usepackage{xcolor}

\begin{document}
\title{Characterizing the bulk and turbulent gas motions in galaxy clusters}
%
%

\author{\firstname{Simon} \lastname{Dupourqué}\inst{1}\fnsep\thanks{\email{sdupourque@irap.omp.eu}} \and 
	\firstname{Etienne} \lastname{Pointecouteau}\inst{1} 
	\and
	\firstname{Nicolas} \lastname{Clerc}\inst{1}
    \and
	\firstname{Dominique} \lastname{Eckert}\inst{2}
}

\institute{IRAP, Université de Toulouse, CNRS, UPS, CNES, Toulouse, France
\and
           Department of Astronomy, University of Geneva, Ch. d’Ecogia 16, CH-1290 Versoix, Switzerland
          }

\abstract{%
The most massive halos of matter in the Universe grow via accretion and merger events throughout cosmic times. These violent processes generate shocks at many scales and induce large-scale bulk and turbulent motions. These processes inject kinetic energy at large scales, which is transported to the viscous dissipation scales, contributing to the overall heating and virialisation of the halo, and acting as a source of non-thermal pressure in the intra-cluster medium. Characterizing the physical properties of these gas motions will help us to better understand the assembly of massive halos, hence the formation and the evolution of these large-scale structures.  We base this characterization on the study of the X-ray and Sunyaev-Zel'dovich effect brightness fluctuations.  Our work relies on three complementary samples covering a wide range of redshifts, masses and dynamical states of clusters. We present the results of our X-ray analysis for the low redshift sample, X-COP, and a subsample of higher redshift clusters. We investigate the derived properties according to the dynamical state of our clusters, and the possibility of a self-similar behaviour based on the reconstructed gas motions power-spectra and the correlation with various morphological indicators.
}
\maketitle
\section{Introduction}
  The baryonic content of galaxy clusters is largely dominated by the hot gas that constitutes the intracluster medium (ICM). The processes governing the growth, as well as the physics of the ICM, introduces perturbations at various scales within the gas. The resulting turbulent motions transport kinetic energy, which cascades to the dissipation scale and contributes to the virialisation of the halo through non-thermal heating of the ICM \citep[e.g.][]{ vazza_turbulent_2018}.
  
  Direct measurements of these turbulent processes can be achieved using spatially resolved X-ray spectroscopy \citep{bohringer_x-ray_2010, the_hitomi_collaboration_quiescent_2016}, and will be fully enabled in the future with the upcoming XRISM and Athena missions \citep{xrism_science_team_science_2020, nandra_hot_2013}. These turbulent processes are expected to induce fluctuations in the thermodynamic properties of the ICM, that should be detectable in the related observables \citep{simionescu_constraining_2019}. For instance, the X-ray surface brightness and Sunyaev-Zel'dovich distortion scale respectively with the squared density and pressure, integrated along the line of sight. Studies on the nearby Coma and Perseus clusters \citep{churazov_x-ray_2012, zhuravleva_gas_2015, khatri_thermal_2016} have demonstrated the feasibility of this approach in both X-ray and SZ, and was already extended to a small sample ($N=10$) to derive statistical trends \citep{zhuravleva_gas_2018}.
  
  With this work, we aim at characterizing the X-ray surface brightness and SZ distortion fluctuations for a large cluster sample ($N>150$) spanning a large range of redshifts, masses and dynamical states. This should allow us to better constrain the properties and impact of  turbulence in the ICM, considering the assembly of massive halos.

\section{Data}

We rely on three samples of clusters built over the Planck \citep{the_planck_collaboration_planck_2014, the_planck_collaboration_planck_2016} and the Atacama Cosmology Telescope \citep[ACT,][]{mallaby-kay_atacama_2021} cluster catalogues:

\begin{itemize}
    \item The XMM Cluster Outskirts Project  \citep[X-COP,][]{eckert_xmm_2017}, a SZ selected sample based on the Planck catalogue, designed to study the outer regions of 12 nearby ($z<0.1$) galaxy clusters.
    \item The Cluster HEritage project with XMM-Newton \citep[CHEX-MATE,][]{the_chex-mate_collaboration_cluster_2021} : a SZ selected sample, based on the Planck catalogue, of 118 clusters designed to obtain an accurate vision of the statistical properties of the cluster population at low to intermediate redshifts ($0.1 < z < 0.5$).
    \item The NIKA2 SZ Large Program \citep[LPSZ@NIKA2,][]{mayet_cluster_2020}, a SZ-selected sample from both Planck and ACT catalogues, including 50 intermediate to high redshift clusters ($0.5 < z < 0.9$) designed to perform evolution studies and cosmological analysis.
\end{itemize}

In this paper, we present our preliminary results on the full X-COP sample, and pilot sub-samples of 8 and 4 clusters for CHEX-MATE and LPSZ@NIKA2, respectively. 

\begin{figure}[t!]
 \centering
 \includegraphics[height=0.47\textwidth]{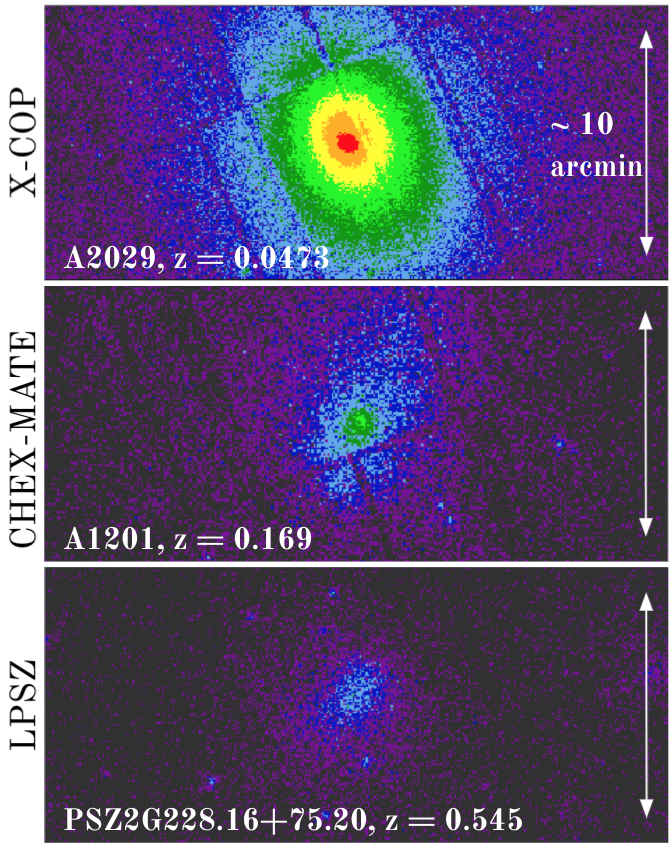}   
 \includegraphics[height=0.47\textwidth]{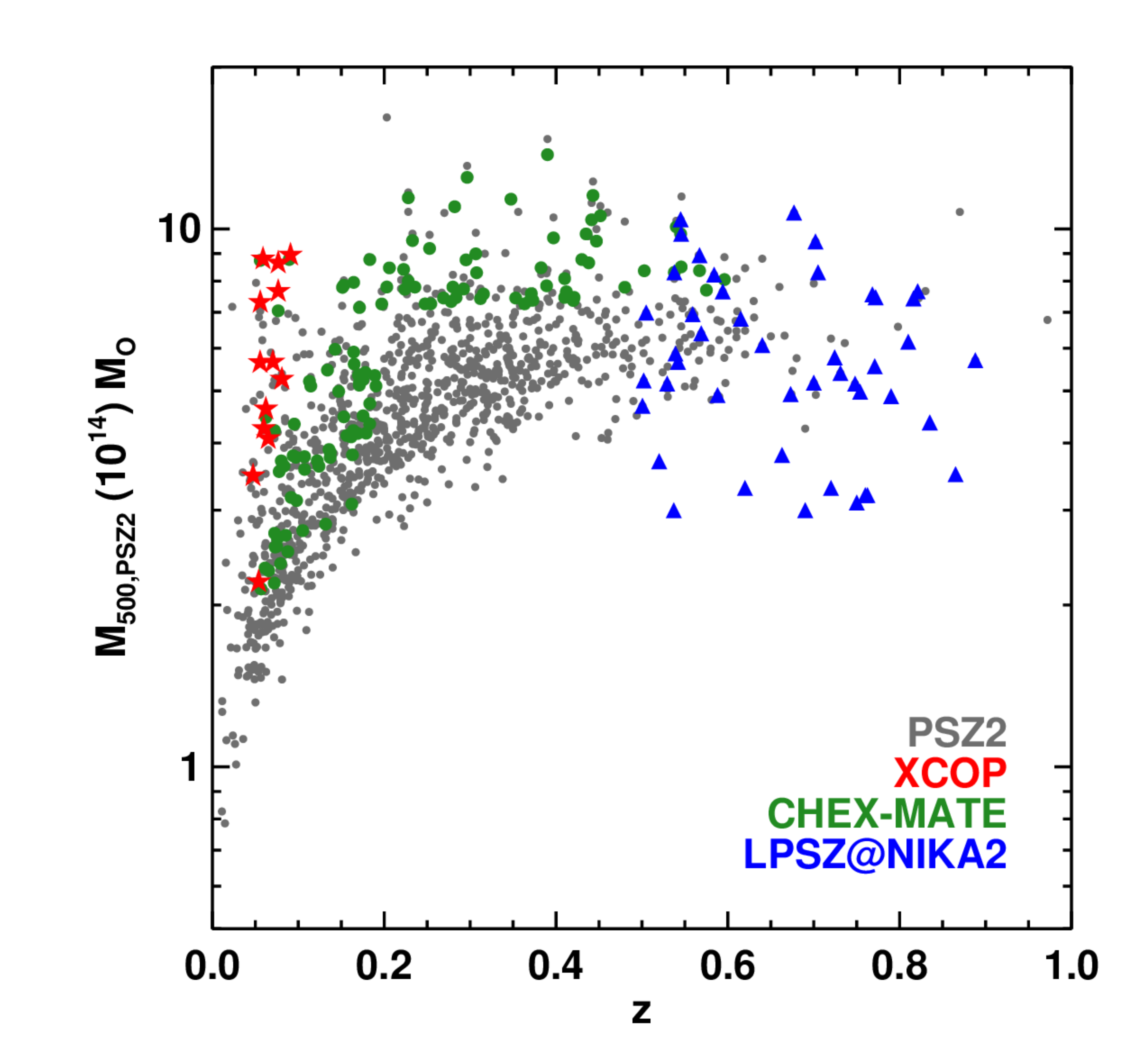}     
  \caption{\textbf{Left:} XMM-Newton image (0.7-1.2 keV band) of A2029, A1201 and PSZ2G228.16+75.20 which belong to X-COP, CHEX-MATE and LPSZ@NIKA2, respectively. \textbf{Right:} Distribution of the X-COP, CHEX-MATE and LPSZ@NIKA2 samples alongside the PSZ2 catalogue, in the mass-redshift plan.}
  \label{fig:plan}
\end{figure}

\section{Method}
\subsection{Mean profile determination}

To determine the mean X-ray surface brightness emission from the cluster, an analytical elliptical model is fitted on the photon count image. We first estimate the centroid of emission and the ellipticity using a principal component analysis in a region of $0.5 R_{500}$, centred on the emission peak. Then we extract the total photon count in concentric elliptical annuli to obtain a radial surface brightness profile with a 10" binning. The parameter distributions of a $\beta$-Model \citep{cavaliere_reprint_1976} fitting this profile is determined using Bayesian inference, assuming that the number of counts in each annulus follows a Poisson distribution.

\subsection{Fluctuation map and 2D power spectrum}

The surface brightness image $I$ is directly dependent on the 3D density profile $n_e$, which can be decomposed into a mean component $n_0$ and relative fluctuations $\delta$: 

\begin{equation}
    I(x,y) = \int^{+\infty}_{-\infty} \Lambda(T) n_e^2(\vec{r})\: dz = \int^{+\infty}_{-\infty} \Lambda(T) n_0^2(\vec{r})(1+\delta(\vec{r}))^2 \: dz
\end{equation}

By linearizing for $\delta$ and dividing by the previously determined mean count image $I_0$, we obtain the map of surface brightness fluctuations, $J$ \citep[see][]{churazov_x-ray_2012}: 

\begin{equation}
    J(x,y) = \frac{I(x,y)}{I_0(x,y)} = 1 + 2 \int^{+\infty}_{-\infty} \eta(\vec{r})\delta(\vec{r}) \: dz; \quad \eta = \frac{n^2_0(\vec{r})}{\int^{+\infty}_{-\infty}  n^2_0(\vec{r}) dz}
\end{equation}

The associated power spectrum, latter referred to as $P_{2D}$, is defined as the squared complex modulus of $\mathcal{F}_{2D}\{ J \}$, the 2D Fourier transform of the fluctuation map. Assuming that the fluctuation field is isotropic and that the emissivity profile is spherically symmetric, the 2D power spectrum is only a function of $k_\rho = ||\vec{k}_\rho||$: 

\begin{equation}
    P_{2D}(k_\rho) = | \mathcal{F}_{2D}\{ J \}(k_\rho)|^2 = \left|\int J(\vec{\rho}) e^{ -2i\pi\vec{k_\rho}.\vec{\rho}}d^2\vec{\rho} \,\right|^2
\end{equation}

For each cluster, the azimuthally averaged power spectrum, $P_{2D}$, is computed following the method introduced by \citep[Arévalo \& al, ][]{arevalo_mexican_2012}. It uses Mexican hat filtering at various scales and handles properly masked data (e.g., source exclusions, gaps, and borders effects). The uncertainties due to the shot noise are estimated by computing $P_{2D}$ for 100 Poisson realizations of the image $I_0$. The power spectrum is also affected by the uncertainty related to the sample variance, which we consider in the following way: assuming that the velocity field is Gaussian, and as in this context, the density fluctuations field is proportional to velocity fluctuations \citep{zhuravleva_relation_2014, gaspari_relation_2014}, the standard deviation of the 3D power spectrum, $P_{3D}$,  is proportional to the power spectrum itself. As the projection operations from $P_{3D}$ to $P_{2D}$ are linear, we assume that the variance of $P_{2D}$ follows the same law. We therefore added to the $P_{2D}$ variance a term of $(P_{2D}/3)^2$.

\subsection{Projection and fitting of the 3D power spectrum}

\begin{figure}[t!]
 \centering
 \includegraphics[width=0.47\textwidth]{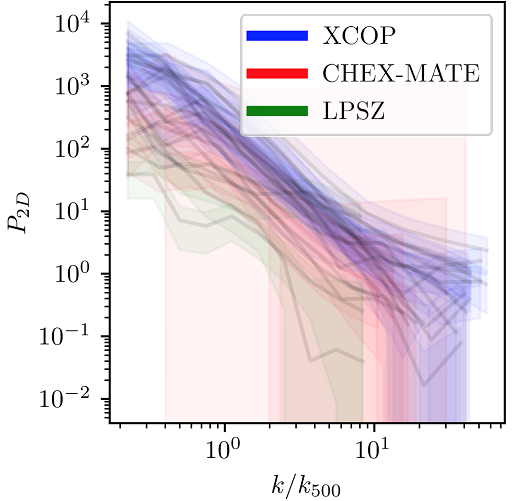}     
 \includegraphics[width=0.48\textwidth]{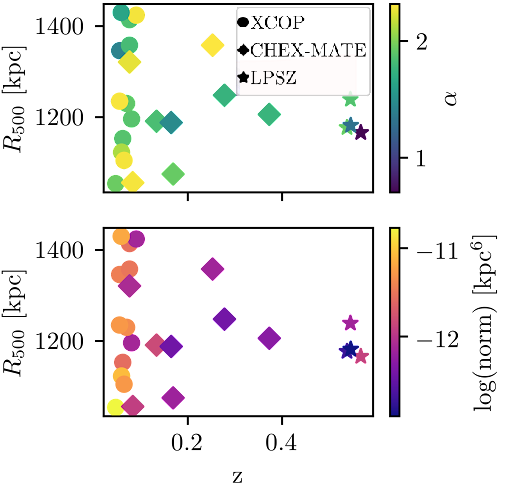}   
  \caption{\textbf{Left:} 2D power spectra, $P_{2D}$ [kpc$^{4}$], for the set of clusters we are using (See. Sec. 2) plotted as a function of the normalized scale, $k/k_{500}$. The shaded envelopes correspond to the $1\sigma$ dispersion about each individual $P_{2D}$. \textbf{Right:} 3D power spectrum best-fit slope $\alpha$ (11/3 corresponds to a Kolmogorov cascade) and normalization in the $R_{500}$-$z$ plane.}
  \label{fig:nada}
\end{figure}

To recover the properties of the inherent 3D density fluctuations power spectrum, we first express the previously obtained 2D power spectrum as a function of the $P_{3D}$. After introducing the power spectrum of the profile $P_\eta$, of the PSF $P_{\text{PSF}}$, the power spectrum of the Poisson noise $P_{\text{Poisson}}$ and by designating the 2D convolution with the $**$ operator, we can show that the observed $P_{2D}$ is linked to the $P_{3D}$ with the following dependence : 

\begin{equation}
\label{eq:projection}
    P_{2D}(k_\rho) =P_{\text{Poisson}} + 4 P_{\text{PSF}}\int \left\{P_\eta ** P_{3D}\right\} (k_\rho, k_z)\:dk_z 
\end{equation}

We use a simple power law representation, $P_{3D} = 10^{A}\left((k_\rho^2 + k_z^2)^{1/2}/k_{\text{ref}}\right)^{-\alpha}$, where the normalization, $10^A$ [kpc$^{6}$], and slope, $\alpha$, are free parameters. The pivot is set to $k_{\text{ref}} = 10^{-3} \text{kpc}^{-1}$. This model is projected using Eq. (\ref{eq:projection}) and fitted on the measured $P_{2D}$ for each cluster.

\section{Preliminary results}
\subsection{Power spectra for a pilot sub-sample}
The 2D power spectra, computed for our pilot sample, are shown in the left of Fig. \ref{fig:nada}, re-scaled to $k_{500} = 1/R_{500}$, and exhibit a strong power-law behaviour in this test sample. The spectra do not seem to vary significantly from one sample to another. Thus, we do not see any trend that could correlate with the redshift (though the statistics of our intermediate and higher redshift sub-samples are very limited). The best values of our fit of $P_{3D}$ are displayed in the $R_{500}-z$ plane in Fig. \ref{fig:nada}, right panel. From this preliminary analysis, we do not see any particular trend of the slope of $P_{3D}$ with the characteristic size or redshift of our clusters, while the normalization decreases with the redshift. This effect will be investigated in a future article. Nevertheless, as the error propagation has not yet been done on these quantities, we cannot draw any solid conclusion about these results.

\subsection{Correlation with morphological indicators}
\begin{figure}[t!]
 \centering
 \includegraphics[width=\textwidth]{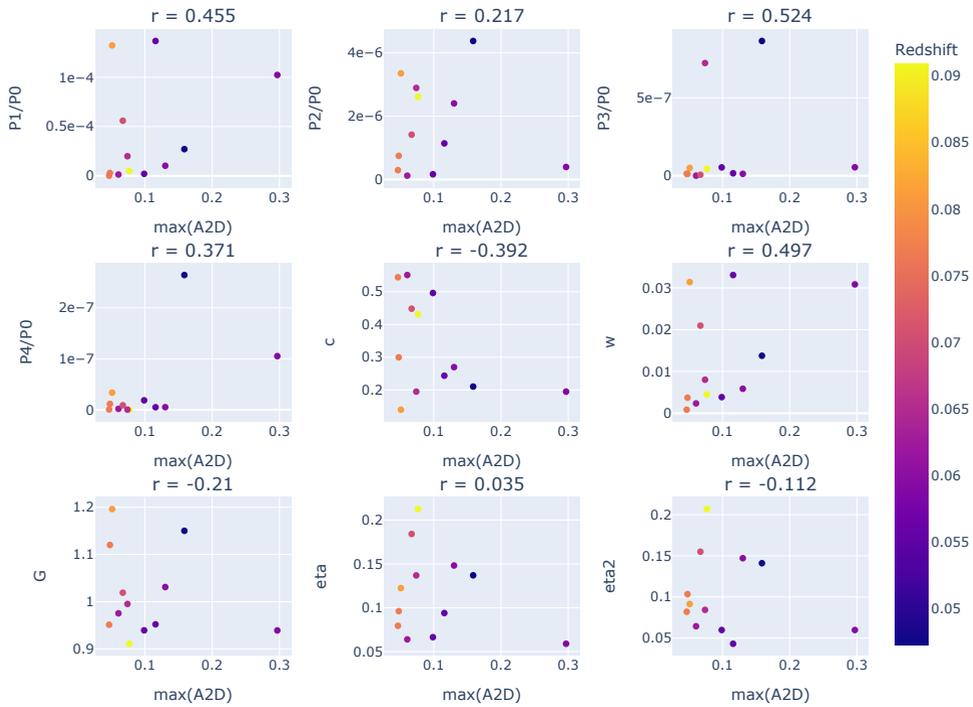}    
  \caption{Morphological indicators w.r.t. the maximum of $A_{2D}$ for the X-COP sample, with the corresponding Spearman coefficient}
  \label{fig:indicators}
\end{figure}

To search for links between the dynamical state of the clusters and the role of turbulence in the ICM, we investigate  the correlation between our 2D-spectra best parametrization and a set of  morphological indicators. We use the following subset of indicators: the centroid shift $w$, the eccentricity $\eta$ (from moments) and $\eta_2$ (from elliptical fitting), the power ratios $P_{i=1-4}/P_0$ , the concentration $c$ ad the Gini coefficient $G$; see  \citep[Lovisari \& al, ][]{lovisari_x-ray_2017}. As the maximum of the characteristic amplitude of the $P_{2D}$ is strongly linked to the 1D Mach number \citep{gaspari_relation_2014}, we compute the Spearman rank  correlation coefficient between the maximum value of $A_{2D} = \sqrt{2\pi k^2 P_{2D}(k)}$ and the various morphological indicator. The Spearman coefficient, $r$, is shown in Fig. \ref{fig:indicators} for the X-COP sample. We observe weak positive correlations for the power-ratios and centroid shifts, which is expected as they reflect the fluctuations at various scales and the disturbances of the surface brightness profile.
\section{Perspectives}

We presented here a preliminary analysis of the power spectrum of brightness fluctuations for a sample of galaxy clusters. We will test models with higher levels of complexity for $P_{3D}$ by including, for instance, the injection scale starting the underlying turbulent cascade. We plan to investigate the correlation with morphological indicators, reflecting the dynamical state of each of our clusters, directly in comparison to the parametrization of  $P_{3D}$, to establish the dependencies between the 3D power spectrum and the dynamical state of the cluster. We will extend our study to the whole CHEX-MATE and LPSZ@NIKA2 samples. We also plan to extend this work to the fluctuations of the SZ signal (directly related to the fluctuations of the ICM pressure), making use of the Planck, ACT and NIKA-2 data. 
%
\bibstyle{woc.bst}
\bibliography{proc.bib}

\end{document}